# Fermat's last theorem, the Riemann hypothesis, and synergistic glasses

J. C. Phillips

*Dept. of Physics and Astronomy, Rutgers University, Piscataway, N. J., 08854-8019*

## Abstract

The synergistic formation of "zero" (exponentially small) resistance states (ESRS) in high mobility two-dimensional electron systems (2DES) in a static magnetic field B and exposed to strong microwave (MW) radiation exhibits hysteresis and an anomalous logarithmic dependence on MW power. These synergistically organized states appear to be associated with a new kind of energy gap $\Delta$.  Here I show that a microscopic quantum model that involves the Prime Number Theorem explains the MW power dependence of the energy gap $\Delta$. The explanation involves remarkable convergence with the most abstract ideas of modern mathematics.

Two experimental groups have discovered giant magnetoresistance oscillations associated with low-field cyclotron resonance in two dimensional electron gases ((2DEG) immersed in a microwave (MW) bath that were previously used to study quantum Hall effects. In very high mobility samples the minima of these oscillations correspond to exponentially small resistance (ESR)[1-3].   The dependence of ESR on MW power P is anomalous[4,5], exhibiting hysteresis similar to that found in many nucleation processes in materials science[6].  Starting from P = 0, there is an incubation regime at low power levels, where the resistance change very slowly with increasing power, followed by a steep decrease in ESR at a higher power levels above $P = P_1$.  In the steeply decreasing (growth or formation) range $P_1 < P < P_2$, the resistivity is linear in lnP.  Starting from the fully formed state at high power $P > P_2$, ESR is maintained to a lower power level P' ~ P/2, and in the range $P_1' < P' < P_2'$ the resistivity is again linear in lnP', but with a slope that is ~ 3 times larger.

The anomalous kinetics involved in the growth or formation of zero resistance states (ZRS) may provide deeper insight into the nature of these states than the positions (oscillatory phase) of the resistance maxima and minima.   The latter have been explained[7] by Zudov in terms of the density of states $N(E)$ of Lorentzian broadened Landau levels

$$N(E) = \Sigma \, f(z_n) \qquad (1)$$

with $f(x) = \Gamma/(x^2 + \Gamma^2)$ and $z_n = E - n\omega_c$.  The phases characteristic of large n are $\varphi = \pm\pi/4$, and these are found from the condition for maximal level mixing,

$$d^2 N(E)/dE^2 = 0 \qquad (2)$$

with $\Gamma \gg x$.  In the opposite limit $\Gamma \ll x$, $\varphi$ becomes linear in n, in agreement with experiment at lower T.

Although (2) properly defines the positions of the oscillatory resistance extrema, it tells us nothing about their magnitude, or the nature of ZRS and the origin of ESR.  Such resistance probably arises because of the formation of an energy gap $\Delta$ that plays the role of order parameter for ZRS.  If the resistivity is given by

$$\rho \sim \exp(-\Delta/T) \qquad (3)$$

and we disregard the additive incubation constants, so that

$$\rho \sim \ln P \qquad (4)$$

then we find that

$$\Delta \sim \ln\ln P \qquad (5)$$

At first sight (5) seems surprising, but it is not so new.  Euler (1737) considered the series $\Sigma n^{-1}$ up to N, which diverges like lnN.  He then noted that the prime numbers p satisfy the relation

$$\Sigma\, n^{-s} = \Pi(1 - p^{-s})^{-1} \qquad (6)$$

for all s > 1, for which the series are convergent.  For s = 1, the series no longer converges, so Euler took logarithms of  (6) to write

$$\Sigma p^{-1} = \text{lnln}\infty = \ln \ln N \qquad (7)$$

(Today much more is known about such series by means of analytic continuation[8,9].)  Next we will argue that (5) and (7) are equivalent.

Starting from P = 0, the states initially are the free-particle Landau states at $E = n\omega_c$ admixed into impurity states for E between the Landau levels.  The Lorentzian broadened model for the density of states corresponds to the series $\Sigma z^{-s}$  with s = 2 and poles displaced from the real axis by $\Gamma$.  The states that dominate ZRS and produce ESR are an exponentially small fraction of the total number of states, and they need not, and should not, correspond to s = 2, that is, to all, or even most, of N(E). Instead they can be expected to correspond to percolative states that avoid the impurities (obstacles responsible for residual resistance) as well as possible to maximize the screening of the MW fields and power. Such states are semi-localized and should lie at the boundary between localized and extended states.  Where is this boundary?

In our previous discussion[9] of the connection between prime numbers and ZRS (ERS), we noted that the series $\Sigma z^{-s}$ defines the Riemann function $\zeta(s)$.  This function exhibits many properties suggestive of an energy gap  for 0 < s < 1, so that the boundary between localized and extended states could be said to fall at s = 1.  If we think of s = 1 as representing the extreme limit of admixing Landau levels to form ZRS, then this identification seems to be not only natural, but also unique.

There are other ways of reaching the conclusion that an extra factor of $n$ is required in the numerator of the series to describe synergistic states. The synergistic energy gap must be proportional to the product of the Landau energy and the microwave energy, but this requires an extra relaxation time factor ($\tau/\hbar$) in the numerator. With $\omega_c \sim 1/n$ and $\omega_c\tau \sim 1$, this extra factor may correspond to an extra factor of $n$. Because it deals directly with the interaction energies responsible for forming the synergistic energy gap, this alternative justification seems to be more reliable than the observation that interlevel step up transition rates scale as $n$. The extra factor $n$ just accounts for coherently generating the $n$th Landau level from the first by multiply scattering through $n$ MW interactions (always step up), and is suggested by gauge invariance and maximal growth rate.

As the MW power P increases above the incubation threshold, it will admix more and more states to form optimized screening states. The number N of such states should be proportional to the number of photons (or the power P at fixed frequency) above threshold (in the absence of heating effects). [Note that threshold or incubation effects correspond to s > 1 in the series; these are convergent and are irrelevant to the divergent term s = 1.] If we replace N by P in (7), then we obtain (5), assuming that the resistivity is limited by equally weighted scattering into states indexed by prime numbers[9]. Previously one argued that synergistic states, because they are generated by interactions with modular fields, can be indexed by (placed in one-to-one correspondence with) integers. Then the coherent (unscattered) ZR states would be indexed by primes. The density of the latter decreases as lnp, and it is this decrease that accounts for the second ln operator in (5) and (7). The experimental results[4,5] for P above the incubation threshold confirm this prediction[9].

Oscillatory magnetoresistance is an example of a modular function[10]. The proof of Fermat's last theorem was stimulated by the Taniyama-Shimura conjecture (1955) concerning a possible relation between modular functions and elliptic curves, two apparently unrelated subjects[11]. A similar (1979) conjecture relates force field constraints and real-space topology of self-organized molecular glasses[12], and it has proved to be most successful in predicting their properties. Modular functions are also related to the Riemann ζ function[8,13].  Possible convergence of these topics was discussed before the recent electronic observations[14].

# References

1.  M. A.  Zudov, R. R. Du, J. A. Simmons, and J. L. Reno, Phys. Rev. B **64**, 201311 (2001).

2.  R. G.  Mani, J. H.  Smet, K.  Von Klitzing, V. Narayanamurthi, W.B. Johnson, and V. Umansky,  Nature **420**, 646 (2002); R. G.  Mani,  W.B. Johnson,  and  V. Narayanamurthi, Bull. Am. Phys. Soc. **48**, 461 (2003).

3.  M.A. Zudov, R.R. Du, L. N. Pfeiffer, and K.W. West, Phys. Rev. Lett. **90**, 046807 (2003).

4.  R. G.  Mani, J. H.  Smet, K.  Von Klitzing, V. Narayanamurthi, W.B. Johnson, and V. Umansky, cond-mat/0306388 (2003).

5.  M. A. Zudov (unpublished).

6.  M. Kessler, W. Dieterich, and A Majhofer, Phys. Rev. B  **67**,  134201  (2003).

7.  M. A. Zudov, cond-mat/0306508 (2003).

8.  H. M. Edwards, *Riemann's Zeta Function* (Dover, Mineola, NY, 1974), p. 1.

9.  J. C. Phillips, Sol. State Comm. (in press).

10. J. Stillwell, *Mathematics and Its History* (Springer, New York, 2002).

11. A. D. Aczel, *Fermat's Last Theorem,* Four Walls Eight Windows, New York, 1996; S. Singh and K A. Ribet, Scien. Am. (11) (1997) 68.

12. J. C. Phillips, J. Non-Cryst. Sol. **34**,153 (1979).

13. J. B. Conrey, Notices Am. Math. Soc. (3), 2003 [http://www.ams.org/notices/200303/200303-toc.html].

14. J. C. Phillips, *Phase Transitions and Self-Organization in Electronic and Molecular Networks* (Kluwer, New York, 2001), p.1.